\documentclass[lettersize,journal]{IEEEtran}
\usepackage{amsmath,amsfonts}
\usepackage{algorithmic}
\usepackage{algorithm}
\usepackage{array}
\usepackage[caption=false,font=normalsize,labelfont=sf,textfont=sf]{subfig}
\usepackage{textcomp}
\usepackage{stfloats}
\usepackage{url}
\usepackage{verbatim}
\usepackage{graphicx}
\usepackage{cite}
\usepackage{xcolor}
\begin{document}

\title{Evaluation of Circular Complex Permeability in Single-Crystal Yttrium Iron Garnet at Cryogenic Temperatures}

\author{
    Junta Igarashi\thanks{Manuscript received XXX XXX, 2025; revised XXX XXX, 2025. This work was supported by New Energy and Industrial Technology Development Organization of Japan (NEDO) Grant (No. JPNP14004), Cross-ministerial Strategic Innovation Promotion Program (SIP), “Promoting Application of Advanced Quantum Technologies to Social Challenges” and JSPS KAKENHI 24K22964 and 24K21737.}%
    \thanks{The authors are with the National Institute of Advanced Industrial Science and Technology, Tsukuba 305-8563, Japan (email: junta.igarashi@aist.go.jp)}, 
    Shota Norimoto, 
    Hiroyuki Kayano,~\IEEEmembership{Member,~IEEE,}, \\
    Noriyoshi Hashimoto,
    Makoto Minohara, 
    Nobu-Hisa Kaneko,~\IEEEmembership{Member,~IEEE,}, \\
    Tomonori Arakawa,~\IEEEmembership{Member,~IEEE}
}

\markboth{Journal of \LaTeX\ Class Files,~Vol.~XX, No.~X, XXX~2025}%
{Shell \MakeLowercase{\textit{et al.}}: A Sample Article Using IEEEtran.cls for IEEE Journals}


\maketitle
\IEEEpubid{\begin{minipage}{\textwidth}\ \\[12pt]
0000--0000/00\$00.00~\copyright~2021 IEEE
\end{minipage}} 
\IEEEpubidadjcol

\begin{abstract}
The operation of superconducting qubits requires a sensitive readout circuit at cryogenic temperatures, driving the demand for cryogenic non-reciprocal microwave components such as circulators. However, evaluating these components at low temperatures presents significant challenges for companies and institutions without specialized measurement systems. In the development of such cryogenic non-reciprocal components, the temperature dependence of ferrite’s magnetic properties is the most critical factor. Therefore, an evaluation technique for accurately assessing these properties at cryogenic temperatures is essential.

In this study, we develop a measurement method to characterize low-loss ferrite materials over a temperature range of 300 K to 2 K. The use of the circularly polarized resonance mode \( \text{TE}_{11n} \) enables the direct estimation of circular complex permeability and the determination of key material parameters, including saturation magnetization and damping constant—both essential for assessing the performance of ferrite materials in circulator applications. To validate the reliability of our measurement method, we selected single-crystal YIG as the test material, as its magnetic properties at cryogenic temperatures are relatively well known. This demonstration confirms that our method is effective for characterizing various low-loss ferrite materials that are potential candidates for compact cryogenic non-reciprocal devices.

\end{abstract}


\section{Introduction}
\IEEEPARstart{Q}{uantum} computers have attracted significant attention from both fundamental and technological viewpoints. Among the most promising candidate is quantum computer based on superconducting qubits, which operate at cryogenic temperatures. Many industries and research institutions are actively working to scale up the number of qubits to advance high-performance quantum computing. To operate superconducting qubits, the connection of microwave circuits between room temperature and cryogenic environment is required. This has led to a growing demand for cryogenic non-reciprocal microwave components, such as circulators and isolators. Circulators and isolators are essential for preventing microwave reflections and ensuring the stable operation of superconducting qubits. Insufficient isolation in these devices can cause unexpected qubit excitation due to thermal noise, degrading overall system performance. While superconducting qubits typically operate at millikelvin temperatures (10–20 mK), nonreciprocal components like circulators are often installed at various stages of the dilution refrigerator (e.g., 50 K, 4 K, 1 K, 100 mK, 20 mK). Therefore, characterizing the temperature dependence of magnetic material properties across a broad temperature range is essential. Although commercial cryogenic circulators (e.g., from QuinStar) are available, their bulky size (40 mm × 40 mm) limits their integration into high-density quantum systems. To overcome this, compact circulators are needed. However, as electromagnetic energy becomes more tightly confined in smaller devices, ferrite materials with ultra-low loss become increasingly critical.

Currently, circulators are designed by first determining the center frequency and then using that value to calculate the saturation magnetization $M_\text{S}$ of the ferrite \cite{marzall2021microstrip}. To achieve the desired $M_\text{S}$, impurities are doped into the ferrite \cite{patton1969effective}. However, certain rare-earth-doped ferrites, such as those containing Gd, Ho, and Dy, exhibit magnetization compensation at low temperatures \cite{pauthenet1958spontaneous}, limiting their applicability for cryogenic circulators. For a ferrite to function as a circulator, it must exhibit a finite difference between the real part of the right- and left-circular complex permeability, which corresponds to a finite magnetization, while maintaining low loss (small imaginary part) in that region \cite{hogan1953ferromagnetic}. Additionally, material properties can change significantly at low temperatures, making it essential to characterize them before device fabrication \cite{konishi1965lumped, arakawa2023calibrated}. However, direct measurement of circular complex permeability at low temperatures is technically challenging, and only a few such measurements exist even at room temperature \cite{artman1955measurement, robbrecht1958measurements}. A recent study Ref. \cite{arakawa2019cavity} demonstrated the excitation of circularly polarized modes in a cavity by evaluating polycrystalline YIG at a single temperature (150 K), but it did not explore the temperature dependence of the magnetic properties. Currently, no well-established experimental framework exists for the cryogenic characterization of ferrite materials across a broad temperature range, posing a major obstacle to the development of low-temperature circulators.

In this study, we aim to establish a measurement method to characterize ferrites from 300 K to 2~K. To this end, we focus on the circularly polarized microwave perturbation method developed by Arakawa \textit{et al}., a powerful technique for non-contact measurements of both complex permeability and complex conductivity~\cite{arakawa2019cavity, arakawa2022microwave}. As a demonstration, we investigate the temperature dependence of the circular complex permeability and key magnetic parameters—namely, saturation magnetization and damping constant—of single-crystal yttrium iron garnet (YIG). Without requiring device fabrication, our approach directly evaluates the magnetic response of YIG under cryogenic conditions. The results confirm that single-crystal YIG can function as a circulator at 2 K. Our methodology provides a valuable tool for screening and optimizing ferrite materials for compact high-performance cryogenic nonreciprocal devices.

\section{Experimental method}
\begin{figure}
    \centering
    \includegraphics[width=0.8\linewidth, trim=0mm 100mm 0mm 0mm,clip]
    {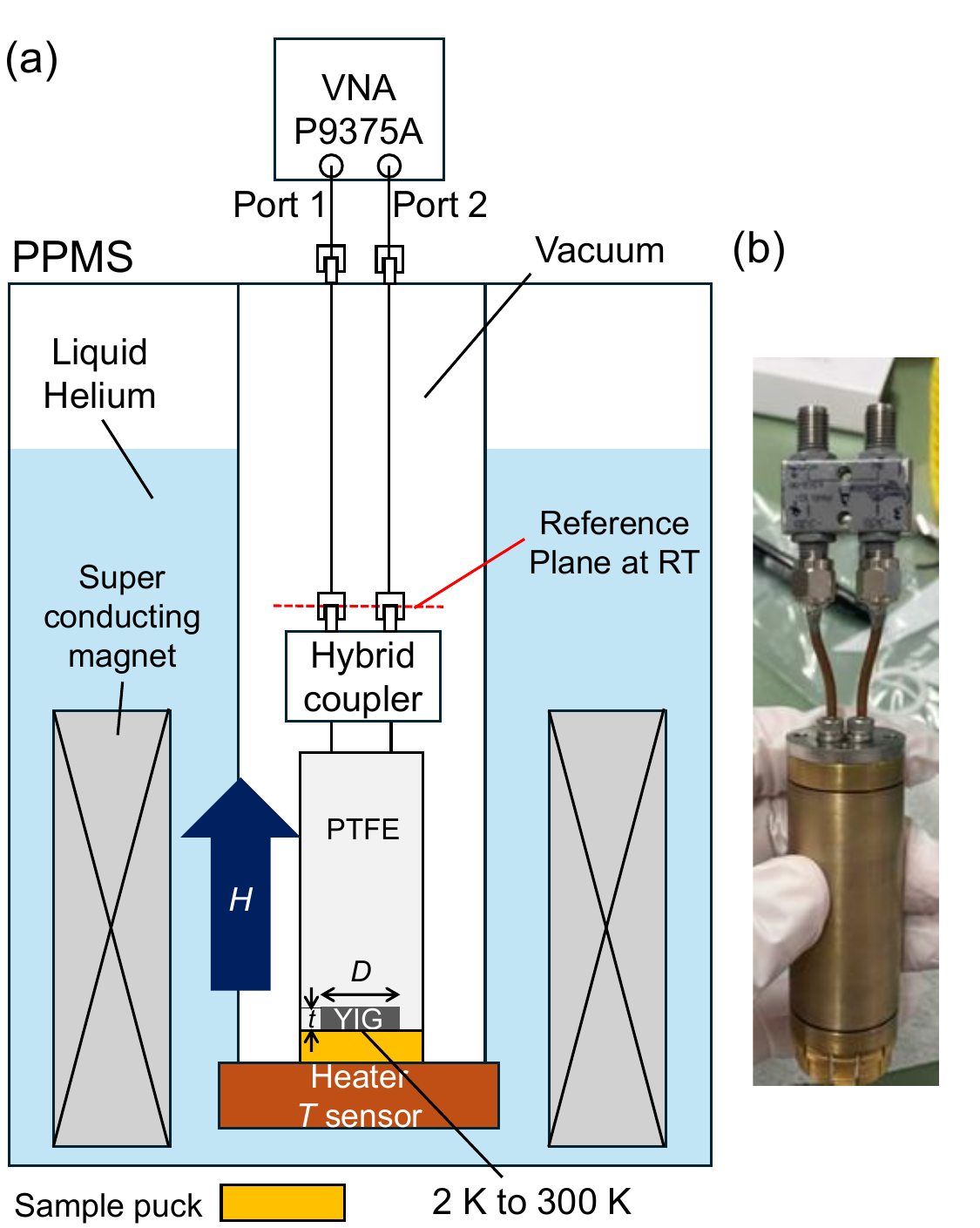}
    \caption{(a) Schematic of the measuremental setup. (b) Photograph of the hybrid coupler and the cavity resotor used in this study.}
    \label{fig1}
\end{figure}

This section describes the experimental methods used in this study. Figure \ref{fig1}(a) shows a schematic illustration of the experimental setup. The circularly polarized microwave is generated by combining two linearly polarized microwave modes with a phase difference of 90 degrees. We employed a cavity resonator and a hybrid coupler to generate the circularly polarized microwave \cite{arakawa2019cavity}.The hybrid coupler (Orient Microwave, BL32-6348-00) is a commercially available component. The cylindrical cavity itself is fabricated from oxygen-free copper (OFC), selected for its high electrical conductivity. The cylindrical cavity resonator was filled with polytetrafluoroethylene (PTFE) and equipped with YIG at its base to facilitate the detection of the multiple transverse electric ($\text{TE}_{11n}$) modes, as shown in Fig. \ref{fig1}(a). The hybrid coupler and cylindrical cavity resonator were connected via SubMiniature version A (SMA) connectors at the end of the RF insert. In this study, the vector network analyzer (VNA) (Keysight P9375A) was utilized for transmission spectroscopy. The hybrid coupler operates by splitting the input microwave signal into two paths with a ±90-degree phase difference, enabling selective excitation of either the right- or left-circularly polarized modes in the cylindrical cavity. The reflected signals are then recombined at the coupler and directed to another measurement port. As a result, the $S_{21}$ and $S_{12}$ parameters correspond to the right- and left-circularly polarized responses, respectively. An electronic calibration kit (Keysight N4691D) was used to calibrate the VNA up to the SMA connector at the tip of the RF insert at room temperature (RT). The reference plane is indicated in Fig.\ref{fig1}(a). While the calibration is inherently temperature dependent, this dependence mainly results in smooth, gradual variations over frequency rather than sharp distortions. The frequency dependence of the background signal at several temperatures is presented in Appendix A. These results show that although the background signal changes with the temperature of the coaxial cable, its effect on the resonance peak is sufficiently small.

\begin{figure}
    \centering
    \includegraphics[width=1\linewidth, trim=0mm 120mm 0mm 0mm,clip]
    {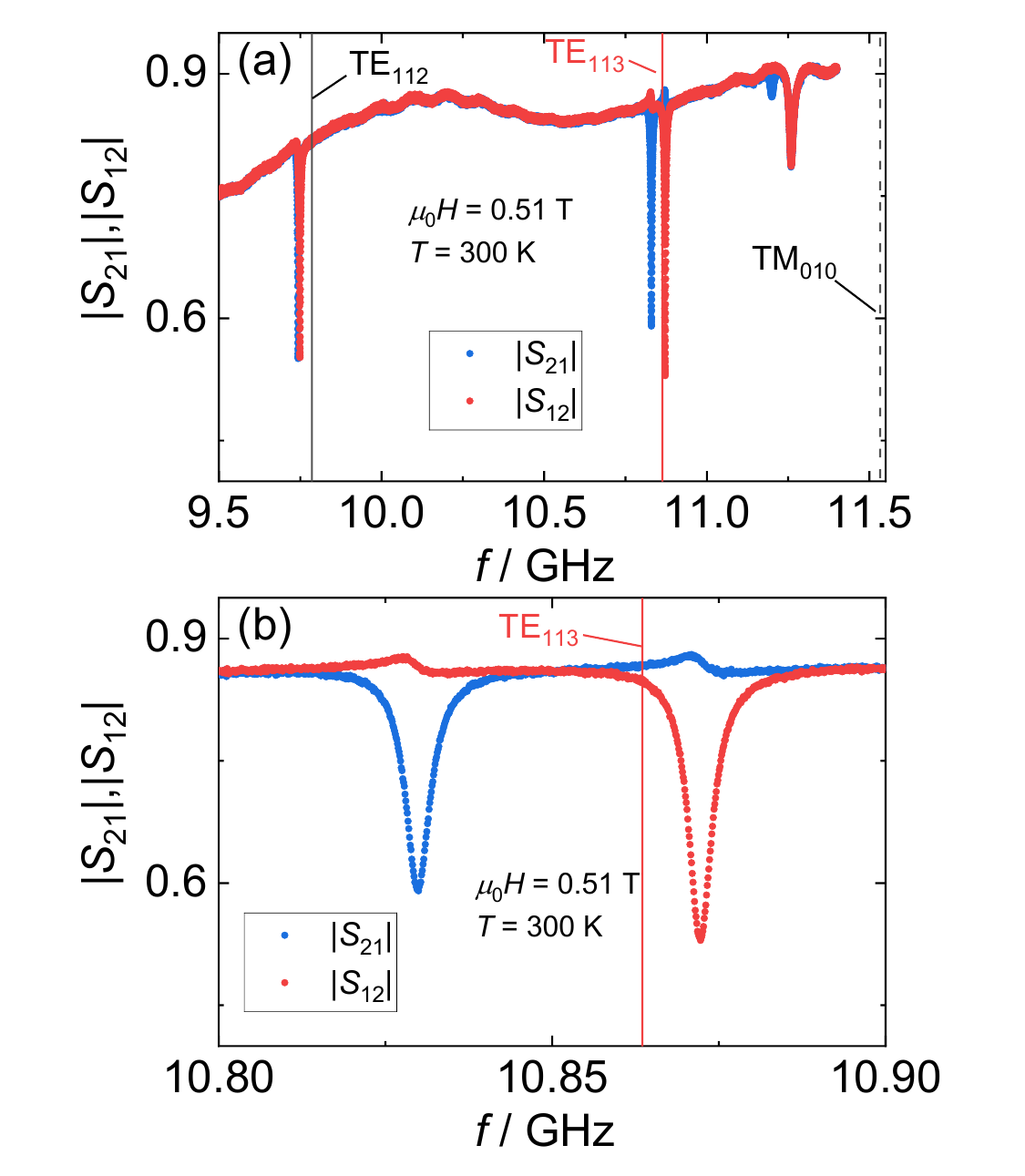}
    \caption{Example of transmission characteristics measured at $T$ = 300 K and $\mu_0H$ = 0.51 T: (a) full-scale view, (b) zoomed-in view. The lines in (a) represent analytically calculated resonance frequencies for each mode using Eqs. \ref{eq:TE11n} and \ref{eq:TM}.}
    \label{fig2}
\end{figure}

The entire setup described above was placed inside a commercially available Physical Property Measurement System (PPMS). Measurements were carried out under perpendicular magnetic fields, ranging from 0 T to 0.8 T at room temperature (RT), and from 0 T to 1 T at low temperatures (LT), with the temperature varied from 2 K to 300 K. The magnetic field generated by the superconducting solenoid magnet in the system is factory-calibrated, ensuring accurate field application. We positioned the cavity within the region where the manufacturer guarantees magnetic field homogeneity. Additionally, the bottom of the cavity is directly connected to the temperature control stage of the PPMS, ensuring good thermal contact. To further improve thermal coupling between the cavity and the YIG sample, we filled any gaps with N grease.
To test the developed method in this study, we selected pure YIG, which does not exhibit magnetization compensation at low temperatures \cite{anderson1964molecular,maier2017temperature}. Furthermore, to minimize microwave loss, we used single-crystal YIG, which has a lower damping constant than its polycrystalline counterpart \cite{sun2012growth, xue2018lowering}.
The YIG samples used in the experiments are disk-shaped with diameters ($D$) and thicknesses ($t$) of 4 mm × 2 mm, 3 mm × 2 mm, 3 mm × 1 mm, and 3 mm × 0.5 mm, respectively. Figure \ref{fig2} shows the measured S-parameters for the YIG sample with 3 mm × 0.5 mm at $T$ = 300 K and $\mu_0H$ = 0.51 T. Scattered symbols correspond to $|S_{21}|$ (blue) and $|S_{12}|$ (red). The lines in Fig. \ref{fig2}(a) indicate the calculated resonance frequencies for the $\text{TE}_{11n}$ and $\text{TM}_{01n}$ modes, obtained using the following equations \cite{pozar2021microwave}:

\begin{equation}
  f_{\text{TE}_{11n}} = \frac{1}{\sqrt{\mu_0\epsilon_{\text{r}}\epsilon_0}} \sqrt{\left(\frac{1.841}{\pi D_\text{cav}}\right)^2+\left(\frac{n}{2L}\right)^2}
\label{eq:TE11n}  
\end{equation}

\begin{equation}
  f_{\text{TM}_{01n}} = \frac{1}{\sqrt{\mu_0\epsilon_{\text{r}}\epsilon_0}} \sqrt{\left(\frac{2.405}{\pi D_\text{cav}}\right)^2+\left(\frac{n}{2L}\right)^2}
\label{eq:TM}  
\end{equation}
where, $\mu_0$ is the permeability of vacuum; $\epsilon_0$ the permittivity of vacuum; $\epsilon_\text{r}$ the relative permittivity ($\epsilon_\text{r}$ = 2.02 for PTFE); $D_\text{cav}$ the diameter of the cavity ($D_\text{cav}$ = 14 mm); and $L$ the length of the cavity ($L$ = 50 mm). The calculated frequencies of each mode generally align with the experimental results, confirming that each mode can be excited as designed. The discrepancy between the experiment and the calculation can be attributed to the fact that the calculation did not account for the YIG sample and the excitation antenna. Figure 2(b) presents a magnified view of the area surrounding the $\text{TE}_{113}$ mode depicted in Figure 2(a). The different resonance frequencies in $|S_{21}|$ and $|S_{12}|$ are confirmed. This indicates that the circularly polarized mode can be selectively excited \cite{arakawa2019cavity}.
\section{Room-temperature experimental results}
\begin{figure*}
    \centering
    \includegraphics[width=0.8\linewidth, trim=0mm 180mm 0mm 0mm,clip]
    {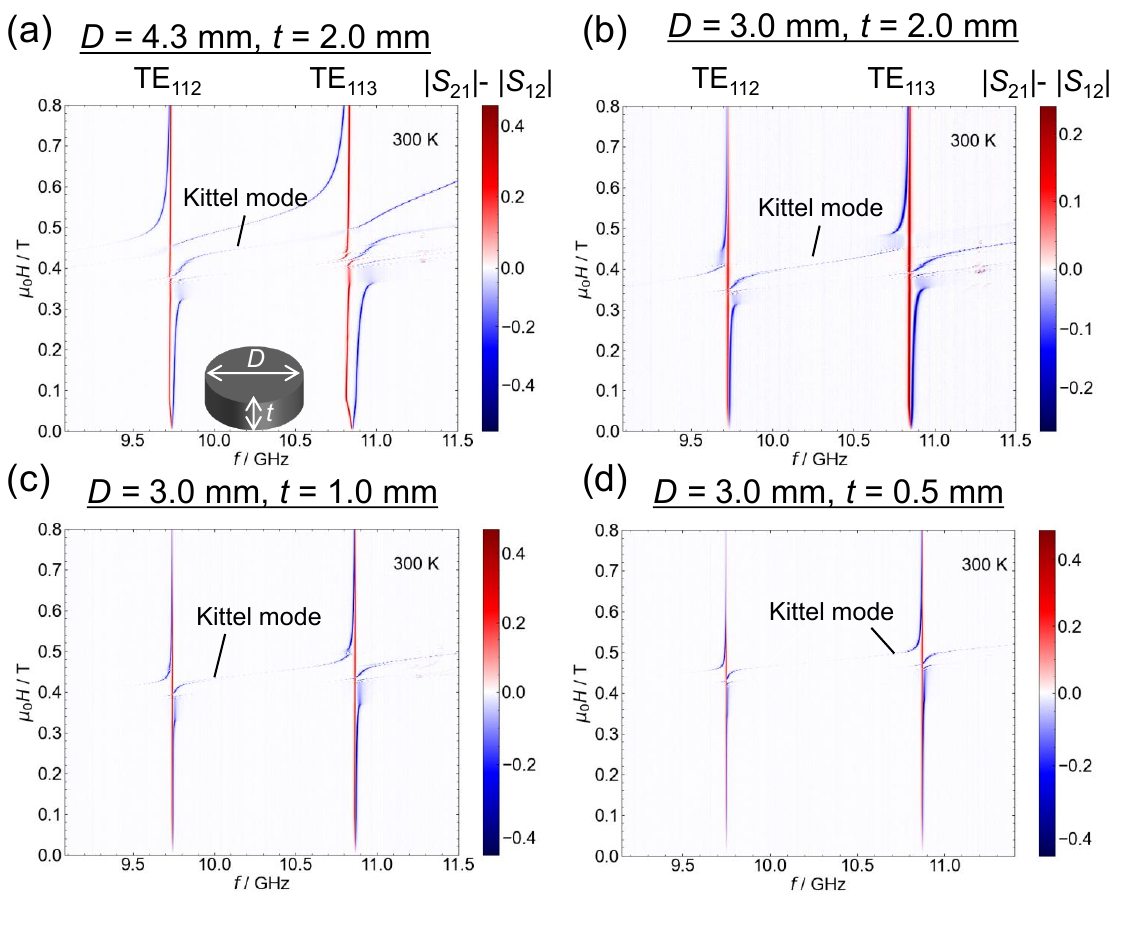}
    \caption{Dependence of $|S_{21}|-|S_{12}|$ on frequency $f$ and applied magnetic field $\mu_0H$ at 300 K for YIG samples with various sizes: (a) $D$ = 4.3 mm, $t$ = 2.0 mm; (b) $D$ = 3.0 mm, $t$ = 2.0 mm; (c) $D$ = 3.0 mm, $t$ = 1.0 mm; (d) $D$ = 3.0 mm, $t$ = 0.5mm. The color bars indicate the amplitude of $|S_{21}|-|S_{12}|$.}
    \label{fig3}
\end{figure*}

In this section, we present the results obtained with YIG samples of varying sizes at RT. Figure \ref{fig3} shows color plots of $|S_{21}|-|S_{12}|$ as a function of frequency ($f$) and applied magnetic field ($\mu_0H$) at RT. We use $|S_{21}|-|S_{12}|$ to eliminate background noise and highlight the right ($|S_{21}|$, blue) and left ($|S_{12}|$, red) circularly polarized modes. As shown in Fig. \ref{fig3}, anticrossings are observed, indicating a strong coupling between the ferromagnetic resonance (FMR) mode, also known as the Kittel mode, and the right circularly polarized mode (blue) \cite{huebl2013high, arakawa2019cavity}. This phenomenon occurs because the magnetization precesses in a counterclockwise direction along the axis of the effective magnetic field, resulting in efficient coupling with the right circularly polarized mode. In contrast, there is no coupling between the Kittel mode and the left circularly polarized mode (red). These results also demonstrate that each circularly polarized mode is selectively excited. Note that the anti-crossing features observed in the red curves of Fig. \ref{fig3}(a) are not due to direct coupling between the left circularly polarized mode and the Kittel mode. Rather, they result from slight symmetry breaking in the cylindrical cavity due to the insertion of the YIG sample and excitation antenna. This imperfection allows a small admixture of the right circularly polarized mode in the left-hand channel, which then couples with the Kittel mode and produces the observed anti-crossing. The coupling strength decreases with the size of the YIG samples, as shown in Fig. \ref{fig3}. This is because the coupling strength is proportional to the square root of the number of spins \cite{imamouglu2009cavity, tabuchi2014hybridizing, zhang2014strongly}. To accurately determine the coupling strength, which is essential for evaluating circularly polarized complex permeability, we selected the smallest YIG sample for further experiments at low temperatures.
\section{Low-Temperature Experimental Results}
In this section, we present the results obtained at low temperatures and demonstrate the temperature dependence of the material parameters derived from our analysis.
Compared to the previous study \cite{arakawa2019cavity}, our analysis method has been extended to enable reliable evaluation of ultra-low-loss ferrite materials, such as single-crystal YIG. This improvement makes it particularly suitable for characterizing candidate materials for cryogenic nonreciprocal devices.

\begin{figure}
    \centering
    \includegraphics[width=1\linewidth, trim=0mm 140mm 0mm 0mm,clip]
    {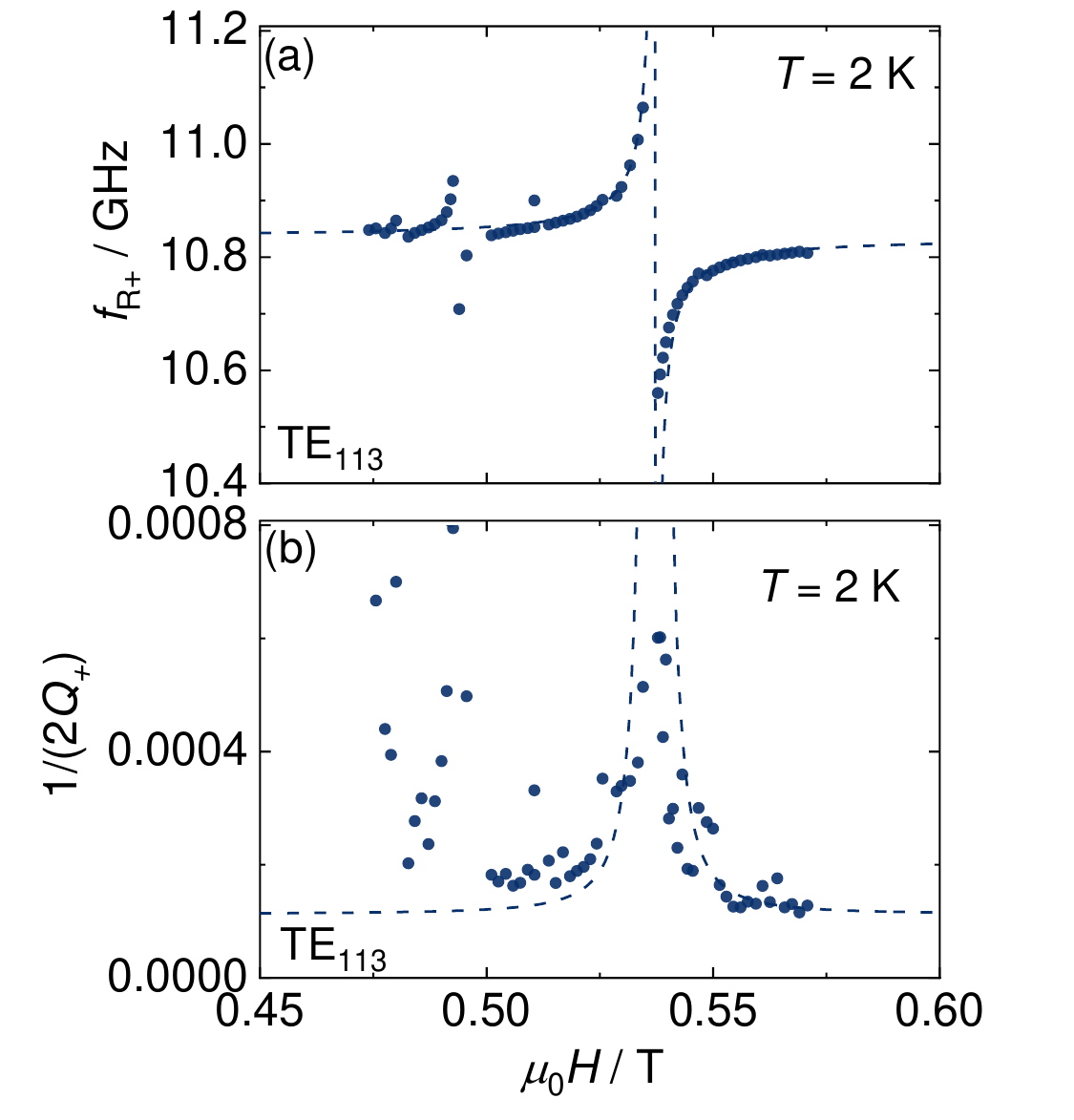}
    \caption{Analysis of the resonance frequency $f_\text{R+}$ and the quality factor $Q_\text{+}$ obtained at $\text{TE}_{113}$ mode. (a) Dependence of the $f_\text{R+}$ on the applied magnetic field $\mu_0H$. (b) Dependence of the 1/(2$Q_\text{+}$) on the applied magnetic field $\mu_0H$. The dotted curves in (a) and (b) represent the fitting curves obtained using Eqs. \ref{eq:fRp} and \ref{eq:Qp}, respectively.}
    \label{fig5}
\end{figure}
In the following analysis, we extract material parameters and determine the coupling strength between the YIG sample and the resonance mode, which is essential for obtaining the complex permeability. This analysis is based on the macrospin model, where all magnetic moments behave uniformly. The dynamic behavior of magnetic moments in magnetic materials is described by the Landau–Lifshitz–Gilbert (LLG) equation, given as follows:
\begin{equation}
    \frac{d\mathbf{M}}{dt} = -\gamma (\mathbf{M} \times \mathbf{H}_{\text{eff}}) + \frac{\alpha}{|\mathbf{M}|} (\mathbf{M} \times \frac{d\mathbf{M}}{dt})
    \label{eq:LLG}
\end{equation}
where, $\gamma$ is the gyromagnetic ratio; $\mathbf{M}$ is the magnetization vector; 
$\mathbf{H_{\text{eff}}}$ the effective magnetic field vector; and $\alpha$ the damping constant. Here, we study the response of the magnetization vector to an effective magnetic field consisting of a circularly polarized field in the x-y plane and a $H_\text{eff}$ in the z direction\cite{artman1955measurement,arakawa2019cavity}. $\mathbf{M}$ and $\mathbf{H_{\text{eff}}}$ are defined as follows:
\begin{equation}
    \mathbf{H}_{\text{eff}} =
\begin{bmatrix}
h_x e^{i2\pi f t} \\
h_y e^{i2\pi f t} \\
H_{\text{eff}}
\end{bmatrix}, \quad
\mathbf{M} =
\begin{bmatrix}
m_x e^{i2\pi ft} \\
m_y e^{i2\pi ft} \\
M_z
\end{bmatrix}
\label{eq:MHeff}
\end{equation}
where, $f$ is the applied microwave frequency; $H_\text{eff} = H_\text{ext}-(N_z-N_x)M_\text{S}$; $N_z$ ($N_x$) the dimensionless demagnetization coefficient along perpedicular (in-plane) to the YIG sample ($N_z > N_x$ for the studied samples); and $M_z$ the magnetization in the z-direction, which is treated as $M_\text{S}$ in this study. 
Also, we define the circularly polarized field $h_{\pm}$ and its response magnetization $m_{\pm}$ as $h_{\pm} = h_x \mp i h_y$ and $m_{\pm} = m_x \mp i m_y$, respectively. The indices $\pm$ represent right ($+$) and left ($-$) circularly polarization. By assuming $|h_{\pm}| \ll H_{\text{eff}}$ and $|m_{\pm}| \ll M_z$, complex magnetic susceptibility is derived\cite{artman1955measurement,robbrecht1958measurements,arakawa2019cavity}:
\begin{equation}
    \begin{aligned}
        \frac{m_{\pm}}{h_{\pm}} &=\frac{\gamma' M_z}{(\gamma' H_{\text{eff}} \mp f) + i \alpha f} \\
        &= \frac{\gamma' M_z (\gamma' H_{\text{eff}} \mp f)}{(\gamma' H_{\text{eff}} \mp f)^2 + (\alpha f)^2} 
        - i \frac{\gamma' M_z \alpha f}{(\gamma' H_{\text{eff}} \mp f)^2 + (\alpha f)^2}
    \end{aligned}
    \label{eq:m/h}
\end{equation}
where, $\gamma'$ is the $\gamma$ divided by $2\pi$ ($\gamma'$ = 28 GHz/T). The real part of Eq. \ref{eq:m/h} is the relative permeability and the imaginary part is the loss. According to the microwave perturbation theory, both components of Eq. \ref{eq:m/h} can be obtained experimentally from the variation of the resonance frequency $f_\text{R}$ and the quality factor $Q$\cite{artman1955measurement}:
\begin{equation}
    \left( \frac{f_{\text{R}\pm} - f_0}{f_0} \right)
    = -a \text{Re} \left[ \frac{m_{\pm}}{h_{\pm}} \right]
    \label{eq:fr_f0}
\end{equation}

\begin{equation}
    \frac{1}{2Q_\pm} - \frac{1}{2Q_0}
    = a \text{Im} \left[ \frac{m_{\pm}}{h_{\pm}} \right]
    \label{eq:1/2Q}
\end{equation}
\begin{figure}
    \centering
    \includegraphics[width=1\linewidth, trim=0mm 60mm 0mm 0mm,clip]
    {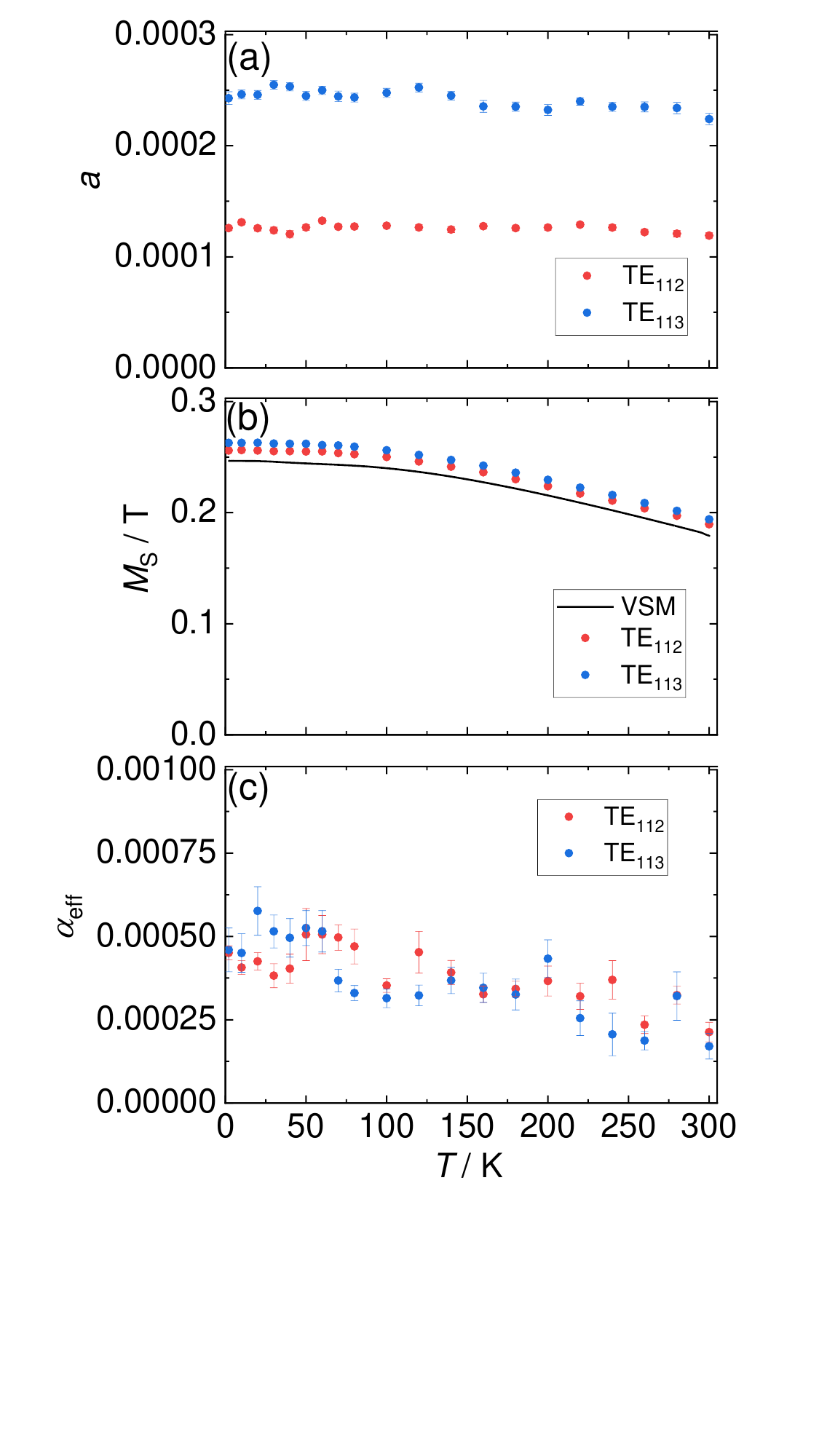}
    \caption{Summary of the obtained fitting parameters as a function of $T$: (a) $a$; (b) $M_\text{S}$; (c) $\alpha_\text{eff}$. The curve in (b) is measured using PPMS with the VSM option. $M_\text{S}$ is measured under a magnetic field of 0.25 T at each temperature point.}
    \label{fig6}
\end{figure}
where, $a$ is the parameter representing the strength of the coupling between the YIG sample and the resonance mode\cite{artman1955measurement,arakawa2019cavity}; $f_0$ the resonance frequency without the sample; and $Q_0$ the quality factor without the sample. In Ref. \cite{arakawa2019cavity}, 0.55 T was used to define $Q_0$ ; however, in our study, this field is already near the resonance condition and thus includes significant influence from the YIG. To avoid this effect, we selected 0 T for $f_0$ and 0.25 T for $Q_0$, which lies sufficiently far from the resonance peak.
Since the magnetic response near the FMR frequency appears only in the right circularly polarized mode, we analyze $|S_{21}|-|S_{12}|$ to eliminate background noise and extract material parameters with higher accuracy. For simplicity, we rewrite Eqs. \ref{eq:fr_f0} and \ref{eq:1/2Q} for the right circularly polarized mode as follows:
\begin{equation}
    f_\text{R+} = f_0 - \frac{a f_0 \gamma' M_\text{S} \left(  \gamma' H_\text{eff} - f_0 \right)}
    {\left( \gamma' H_\text{eff} - f_0 \right)^2 + (\alpha_{\text{eff}} f_0)^2}
    \label{eq:fRp}
\end{equation}

\begin{equation}
    \frac{1}{2Q_\text{+}} = \frac{\Delta f_0}{2 f_0} + \frac{a f_0 \gamma' M_\text{S} \alpha_{\text{eff}}}
    {\left( \gamma' H_\text{eff} - f_0 \right)^2 + (\alpha_{\text{eff}} f_0)^2}
    \label{eq:Qp}
\end{equation}
where, $\Delta f_0$ is the half width at $\mu_0H$ = 0.25 T; and $\alpha_\text{eff}$ the effective damping constant, which will be discussed later. Experimentally, $Q_\text{+}$ is obtained by dividing $f_\text{R+}$ by its the half width. Figure \ref{fig5} shows $f_\text{R+}$ and 1/(2$Q_\text{+}$) as a function of $\mu_0H$ at $T$ = 2 K. The symbol corresponds to the experimental results. To apply Eqs. \ref{eq:fRp} and \ref{eq:Qp} for the analysis, we correct the magnetic field by subtracting $\frac{h}{g \mu_\text{B}} (f_0 - f_{\text{R}+})$, where, $g$ is the $g$-factor ($g$ = 2 for YIG); $\mu_{\text{B}}$ the Bohr magneton; and $h$ the Planck constant. Unlike a previous study \cite{arakawa2019cavity}, to minimize the number of fitting parameters, we numerically calculate $N_z - N_x = 0.57$ for $D = 3.0$ mm and $t = 0.5$ mm using the following equations \cite{beleggia2006equivalent}:
\begin{equation}
    N_z (\tau) = 1 + \frac{4}{3\pi \tau} \left[ 1 - \frac{1}{\kappa} \left( (1 - \tau^2) E(\kappa^2) + \tau^2 K(\kappa^2) \right) \right],
    \label{eq:Nz}
\end{equation}
\begin{equation}
    N_x = \frac{1}{2}(1-N_z),
    \label{eq:Nx}
\end{equation}
\begin{figure*}
    \centering
    \includegraphics[width=0.8\linewidth, trim=0mm 210mm 0mm 0mm,clip]
    {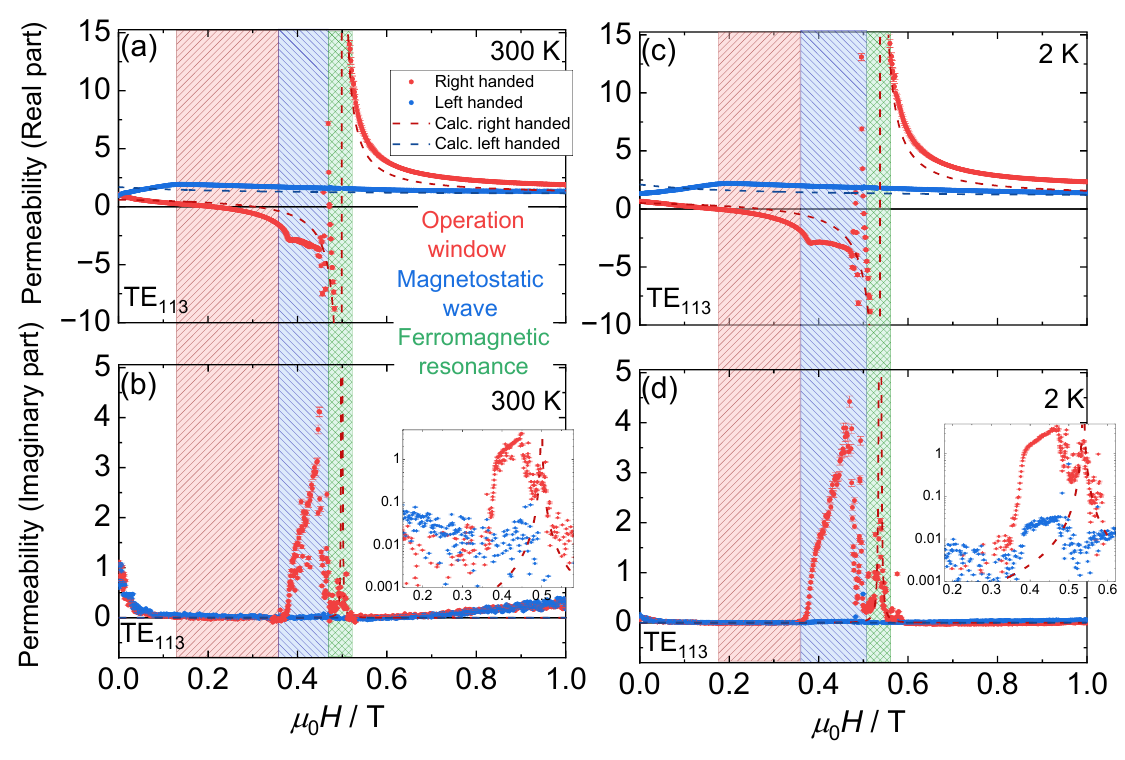}
    \caption{Circular complex permeability: (a) real part at 300 K, (b) imaginary part at 300 K, (c) real part at 2 K, and (d) imaginary part at 2 K. Three regions are highlighted: the operation window (red), where a finite difference exists between the real parts of the right- and left-circular permeability while the imaginary part remains zero; the magnetostatic wave region (blue); and the ferromagnetic resonance region (green). Dotted curves represent calculations based on the macrospin model. The insets in (b) and (d) show the same data plotted with a logarithmic scale on the vertical axis.}
    \label{fig7}
\end{figure*}
where $\tau$ is the aspect ratio of the YIG sample ($\tau = \frac{t}{D}$), $\kappa = \frac{1}{\sqrt{1+\tau^2}}$, $K$ is the type I complete elliptic integral, and $E$ is the type II complete elliptic integral. We fit the Eqs. \ref{eq:fRp} and \ref{eq:Qp} to the experimental results using fitting parameters $a$, $M_\text{S}$, and $\alpha_\text{eff}$. Since it was difficult to determine $a$ and $\alpha_\text{eff}$ simultaneously due to the narrow linewidth in the Fig. \ref{fig5}(b), we first fit Eq. \ref{eq:fRp} to the results for $f_\text{R+}$, assuming that $\alpha_\text{eff}$ is negligible. This result was fitted within the range of $\mu_0H$ corresponding to the $f_0 \pm 100$ MHz frequency range to obtain the values of $a$ and $M_\text{S}$. Then, with $a$ and $M_\text{S}$ fixed, we fit Eq. \ref{eq:Qp} to the experimental results over the same $\mu_0H$ range to determine $\alpha_\text{eff}$. As can be seen in Fig. \ref{fig5}, experimental results are reproduced well by Eqs. \ref{eq:fRp} and \ref{eq:Qp}. The deviation from fitting curve at near $\mu_0H$ = 0.49 T arise from magnetostatic wave modes. arises from magnetostatic wave modes. Our fitting model is based on the macrospin approximation, which assumes uniform spin precession. While this accurately captures the Kittel mode, it does not account for the non-uniform spin dynamics associated with magnetostatic waves. We summarize obtained fitting parameters with respect to $T$ for $\text{TE}_{112}$ and $\text{TE}_{113}$ modes in Fig. \ref{fig6}.

We find that the $a$ remains nearly constant regardless of $T$, suggesting that once the $a$ is determined at one temperature, the same value can be used to obtain the complex permeability at other temperatures. 

In Fig.~\ref{fig6}, to confirm the validity of the presented method, we compare the values of $M_\text{S}$ measured using the Vibrating Sample Magnetometer (VSM) option of the PPMS with those obtained using our method (see Appendix for details on the magnetization measurement). Although the value of $M_\text{S}$ obtained using this method ($\text{TE}_{113}$ mode) is approximately 7 \% larger than that obtained with the VSM, the qualitative trends with respect to temperature variations remain consistent. Moreover, the $M_\text{S}$ values obtained by both methods are generally agree with those reported in a previous study \cite{anderson1964molecular,maier2017temperature}. For instance, Ref.~\cite{maier2017temperature} reports that the saturation magnetization $M_\text{S}$ of YIG is approximately 0.18 T at 300 K and increases to 0.25 T at 5 K. Our results—regardless of the specific measurement method—show that the magnetic properties of single-crystal YIG saturate below 50 K, indicating its suitability for stable circulator operation across the 50 K to 2 K range. Moreover, based on the observed temperature dependence of the saturation magnetization, we can reasonably extrapolate that the magnetic properties remain essentially unchanged even below 2 K. While the exact reason for the discrepancy between the two methods is unclear, one possible explanation is that the symmetry of the cylindrical cavity resonator is slightly disrupted by the excitation antenna and the YIG sample. 

Regarding $\alpha_\text{eff}$, the values obtained for both $\text{TE}_{112}$ and $\text{TE}_{113}$ modes increase as $T$ decreases. This trend could be attributed to trace impurities, such as rare-earth elements, which are known to enhance magnetic relaxation at low temperatures \cite{spencer1959low,dillon1959effects}. Another possible factor is conductor loss. Although losses unrelated to the magnetic material under a uniform magnetic field, are expected to be subtracted in Eq. \ref{eq:Qp}, conductor loss from the Cu layer beneath the YIG—where complex magnetic dynamics occur during ferromagnetic resonance—may still be included in the evaluation. This could lead to an overestimation of the damping constant. For this reason, we use the term $\alpha_\text{eff}$ instead of $\alpha$, which is a material-dependent parameter. This issue arises because the damping constant of the YIG sample studied here is an order of magnitude lower than in a previous study ($\alpha$ = 0.0023)\cite{arakawa2019cavity}, resulting in the very narrow linewidth observed in Fig. \ref{fig5}(b). In addition, Ref.\cite{maier2017temperature} reports that the $\alpha$ of a YIG sphere is approximately $4 \times 10^{-5}$ at room temperature and decreases to about $1 \times 10^{-5}$ at 100 K. To more accurately evaluate the such low intrinsic damping constant, a high-$Q$ resonator such as a dielectric resonator would be required~\cite{roppongi2024microwave}.

\section{Complex Permeability Analysis}

Here, we discuss the circular complex permeability of single-crystal YIG. We obtain the circular complex permeability from the measured \( f_\text{R} \pm \) and \( Q_\pm \) using Eqs. \ref{eq:fr_f0} and \ref{eq:1/2Q}. Since the real part of \( \frac{m_{\pm}}{h_{\pm}} \) represents the relative permeability, the real part of the complex permeability is obtained by adding 1 \cite{arakawa2019cavity}. Figure \ref{fig7} shows the circular complex permeability of single-crystal YIG obtained from the \( \text{TE}_{113} \) mode at 300 K and 2 K. The dotted curves represent calculations using Eqs. \ref{eq:fr_f0}, \ref{eq:1/2Q}, \ref{eq:fRp}, and \ref{eq:Qp} with the obtained fitting parameters. In Fig. \ref{fig7}, we highlight three regions: the operation window (red), where there is a finite difference between the real parts of the right- and left-circular permeability while the imaginary part remains zero; the magnetostatic wave region (blue) \cite{walker1957magnetostatic}; and the ferromagnetic resonance region (green). The significant microwave loss observed in Fig. \ref{fig7} originates from magnetic loss within the YIG, specifically due to excitations of the ferromagnetic resonance mode (green region) and magnetostatic wave modes (blue region). These two types of excitations differ in whether the spins in the YIG precess coherently (Kittel mode) or non-uniform (magnetostatic modes) but result in intrinsic loss within the material. Therefore, to ensure low-loss operation in actual circulator applications, these regions of resonance should be avoided. As shown in Fig. \ref{fig7}, the operation window exists at 2 K, indicating that this YIG sample can function as a circulator at 2 K from a material perspective. The operation window at 2 K is slightly narrower than that at 300 K, which can be attributed to the higher magnetic field required to saturate the magnetization at 2 K (see Appendix for more details). Although the operation window is defined here as the magnetic field range above which the magnetization saturates \cite{arakawa2019cavity}, the circulator should also function at lower magnetic fields if there is no microwave loss. In fact, a previous study experimentally and computationally confirmed that the circulator can operate even with unsaturated magnetization in the ferrite \cite{marzall2021microstrip}. At 2 K, the microwave loss in the low magnetic field region is smaller than at 300 K. As a result, the effective operation window at 2~K is wider than that at 300~K. Regarding the microwave loss, as shown in Figs.~\ref{fig7}(b) and \ref{fig7}(d), for the YIG sample with a low damping constant on the order of $10^{-4}$, the microwave loss in the magnetostatic wave region is significantly larger and extends over a broader range compared to the ferromagnetic resonance region. 
At 2~K, the broadening of the magnetostatic wave region is attributed to the increase in saturation magnetization at lower temperatures. Magnetostatic waves are excited in the field range where the real part of the permeability becomes negative. As the saturation magnetization increases, the splitting between the right- and left-circularly polarized permeabilities becomes larger, thereby expanding the magnetostatic wave region.
It is important to note that the model used for the calculation in Fig.~\ref{fig7} is based on the macrospin approximation and does not incorporate magnetostatic wave contributions. Consequently, the calculated operation window is likely overestimated. This underscores the importance of directly measuring the circular complex permeability. In this work, we employed the circularly polarized microwave perturbation method to demonstrate that single-crystal YIG can effectively function as a circulator at 2~K from a material perspective. Furthermore, we highlight that in ultra-low-loss materials such as single-crystal YIG, microwave loss originating from magnetostatic waves can become significantly more dominant than that from ferromagnetic resonance.

\section{Conclusion}
We have established a measurement method to evaluate the circular complex permeability of single-crystal YIG at cryogenic temperatures down to 2~K using the circularly polarized microwave perturbation method. This parameter is essential for the development of cryogenic circulators. Our results demonstrate that single-crystal YIG can function effectively as a circulator at 2~K. The proposed method allows direct assessment of a ferrite’s suitability for circulator applications without requiring device fabrication, over a wide temperature range from 300~K to 2~K. This approach offers a powerful tool for evaluating ferrite materials under cryogenic conditions and contributes to the development of compact cryogenic circulators and isolators. 

Although the minimum temperature in this study was limited to 2~K due to the constraints of the measurement system, applying this method to a dilution refrigerator setup would enable evaluations down to the millikelvin range~\cite{tabuchi2014hybridizing}. Moreover, based on the observed temperature dependence of the saturation magnetization, we anticipate that the magnetic properties of single-crystal YIG remain nearly unchanged even below 2 K, further supporting its potential as a promising material candidate for millikelvin applications.

\section*{Acknowledgments}
The authors thank J. Kato and S. Ishida for their technical support, M. Roppongi for fruitful discussion, and Y. Yoshida for providing single-crystal YIG.

\appendices

\section{Transmission Characteristics Measured at Various Temperatures}
\begin{figure}[h]
    \centering
    \includegraphics[width=0.8\linewidth, trim=0mm 120mm 0mm 0mm,clip]{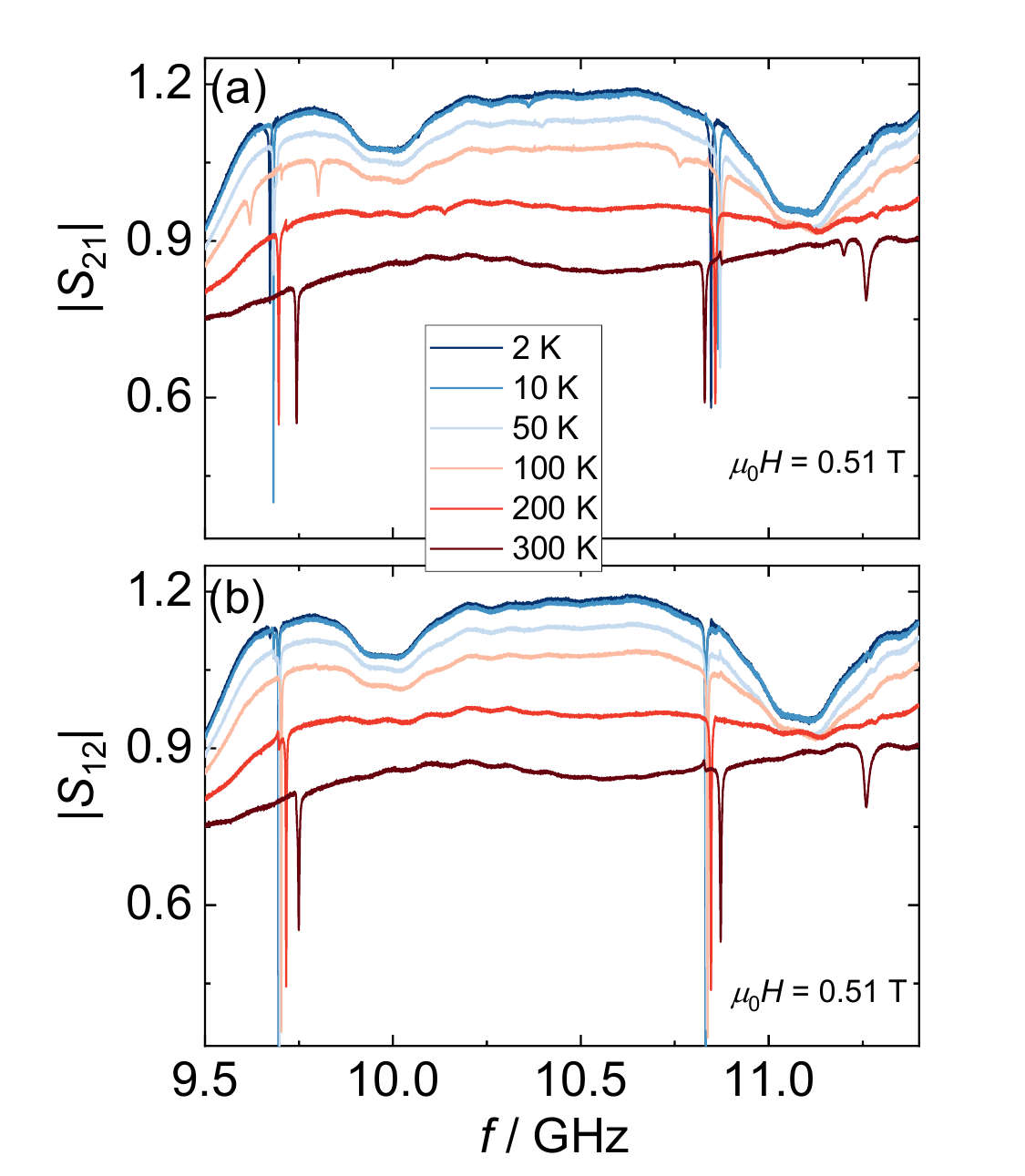}
    \caption{Frequency dependence of the transmission characteristics at several temperatures at $\mu_0H = 0.51$ T: (a) $S_{21}$; (b) $S_{12}$.}
    \label{fig7R}
\end{figure}

Here, we present the results of transmission characteristics measured at various temperatures. Figure~\ref{fig7R} shows the frequency dependence of the background signal at several temperatures. These results show that while the background signal varies with the temperature of the coaxial cable, its effect on the resonance peak remains sufficiently small.

\section{Magnetization Measurement}
\begin{figure}[h]
    \centering
    \includegraphics[width=1\linewidth, trim=0mm 220mm 0mm 0mm,clip]{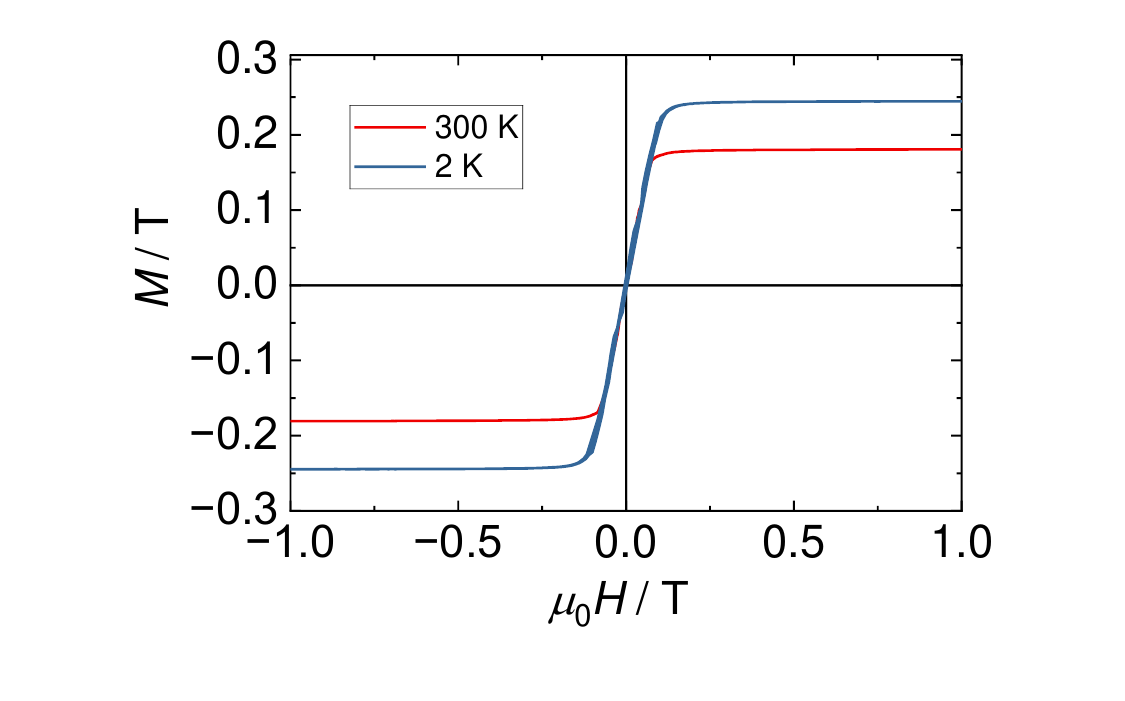}
    \caption{$M$-$H$ curves measured by VSM at 300 K and 2 K.}
    \label{fig8}
\end{figure}

Here, we present the results of magnetization measurements using a PPMS with the VSM option. The YIG ferrite disk used in this experiment has a diameter of 4.3 mm and a thickness of 2.0 mm. Figure~\ref{fig8} shows the magnetization curves measured at 300 K and 2 K. As the saturation magnetization $M_\text{S}$ increases with decreasing temperature, the magnetic field required for saturation at 2 K is higher than that at 300 K. Consequently, the operational window at 2 K is narrower than that at 300 K, as discussed in the main text (see Fig.~\ref{fig7}). Additionally, we measured the temperature dependence of $M$ by increasing $T$ from 2 K to 300 K while recording $M$ at $\mu_0H = 0.25$ T (see Fig.~\ref{fig6}). The results are in good agreement with previous studies~\cite{anderson1964molecular,maier2017temperature}.

\nocite{*}
\bibliographystyle{IEEEtran}
\bibliography{YIG}

\newpage

\begin{IEEEbiography}[{\includegraphics[width=1in,height=1.25in,clip,keepaspectratio]{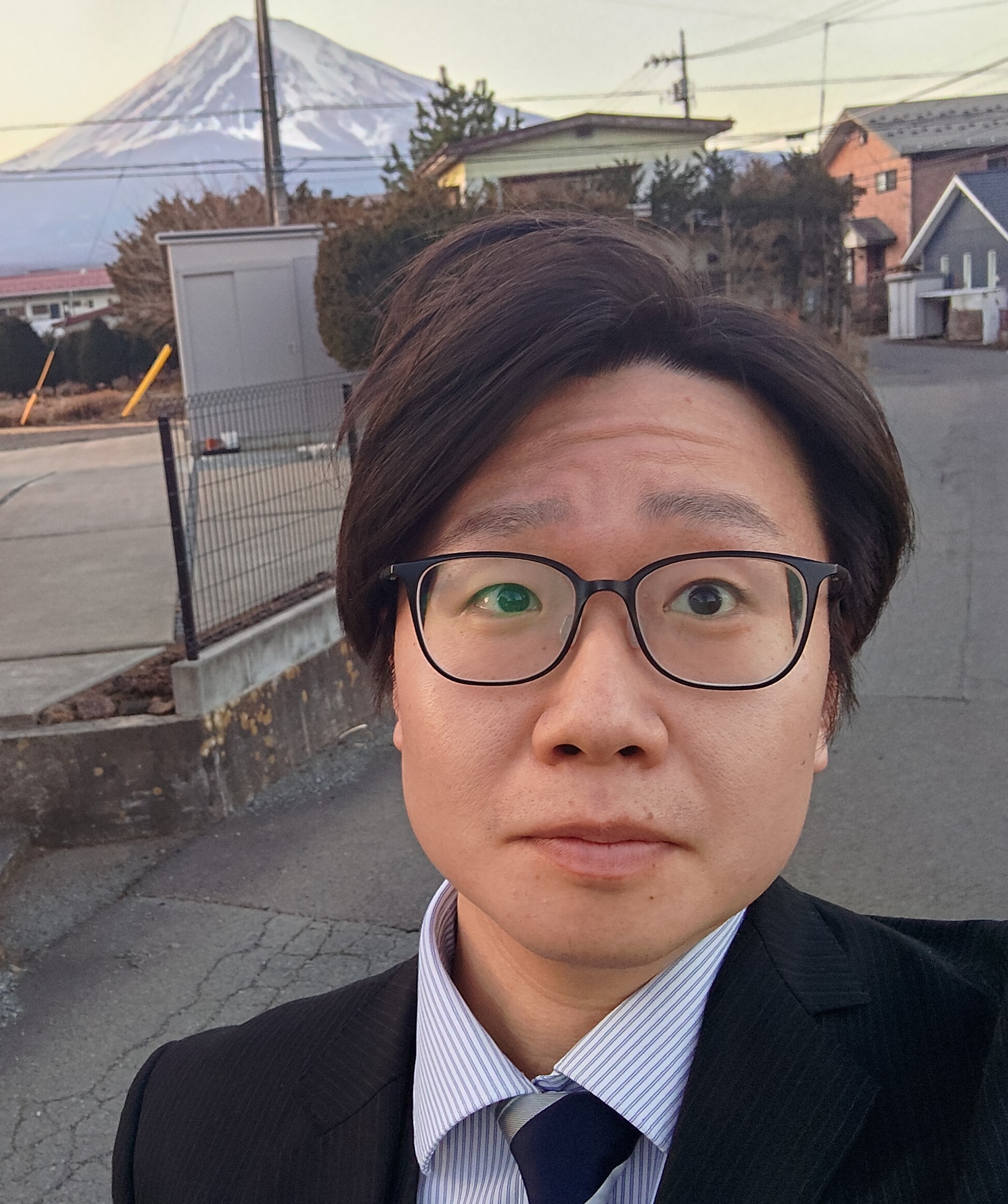}}]{Junta Igarashi}
received his Ph.D. from Tohoku University, Sendai, Japan, in 2021. After a short postdoctoral position at Tohoku University, he joined the University of Lorraine in France as a postdoctoral researcher. In 2024, he joined the Global Research and Development Center for Business by Quantum-AI Technology (G-QuAT) at the National Institute of Advanced Industrial Science and Technology (AIST) as a Researcher. His work at AIST focuses on evaluating microwave components and materials at cryogenic temperatures.
\end{IEEEbiography}

\begin{IEEEbiography}[{\includegraphics[width=1in,height=1.25in,clip,keepaspectratio]{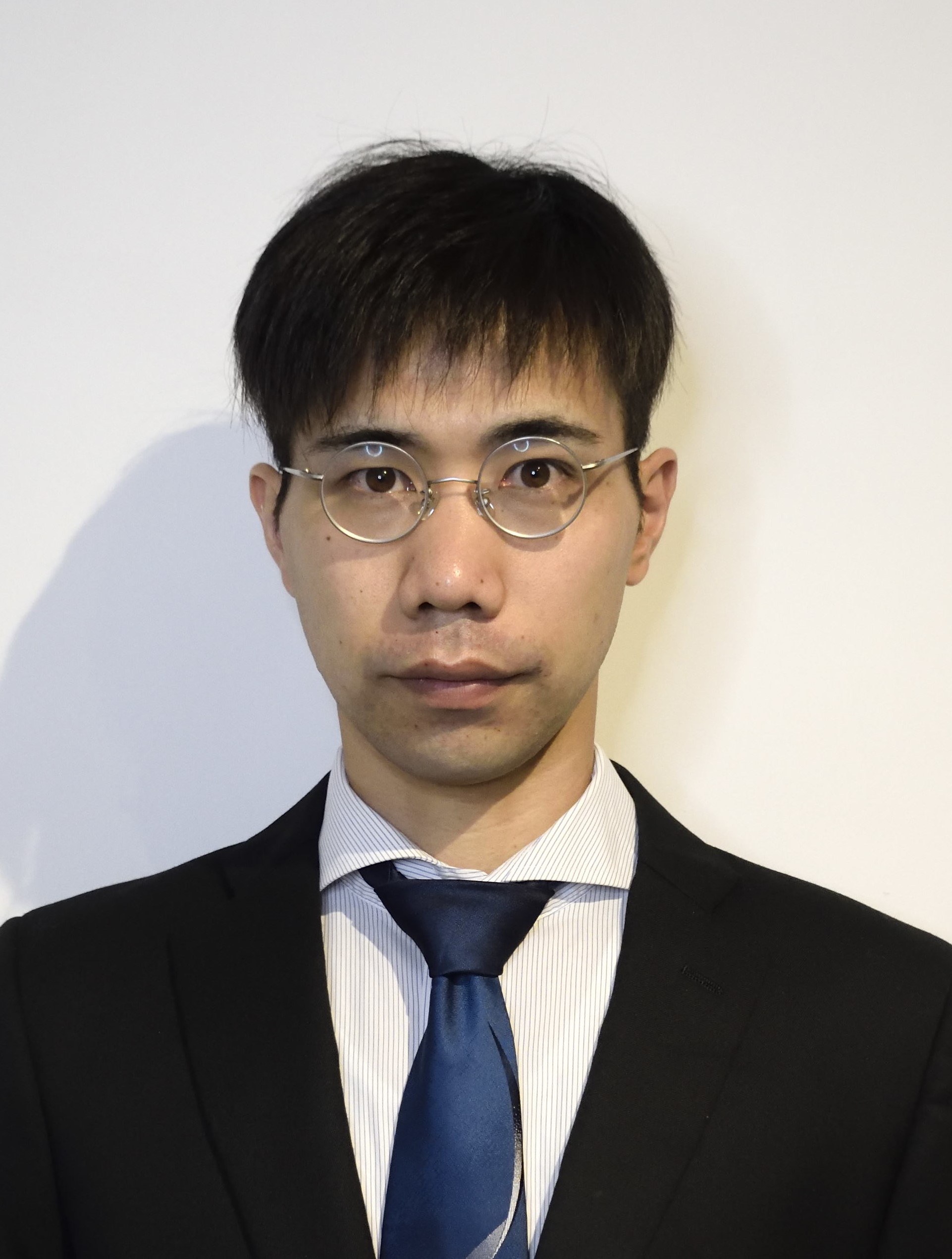}}]{Shota Norimoto}
received his Ph.D. from Osaka University, Toyonaka, Japan, in 2019. He held the position of Higher researcher at National Physical Laboratory (NPL) in the United Kingdom for two years before joining the Global Research and Development Center for Business by Quantum-AI Technology (G-QuAT) at the National Institute of Advanced Industrial Science and Technology (AIST) as a Researcher in 2024. His work at AIST focuses thermal conductance evaluation for microwave components at cryogenic temperatures.
\end{IEEEbiography}

\begin{IEEEbiography}[{\includegraphics[width=1in,height=1.25in,clip,keepaspectratio]{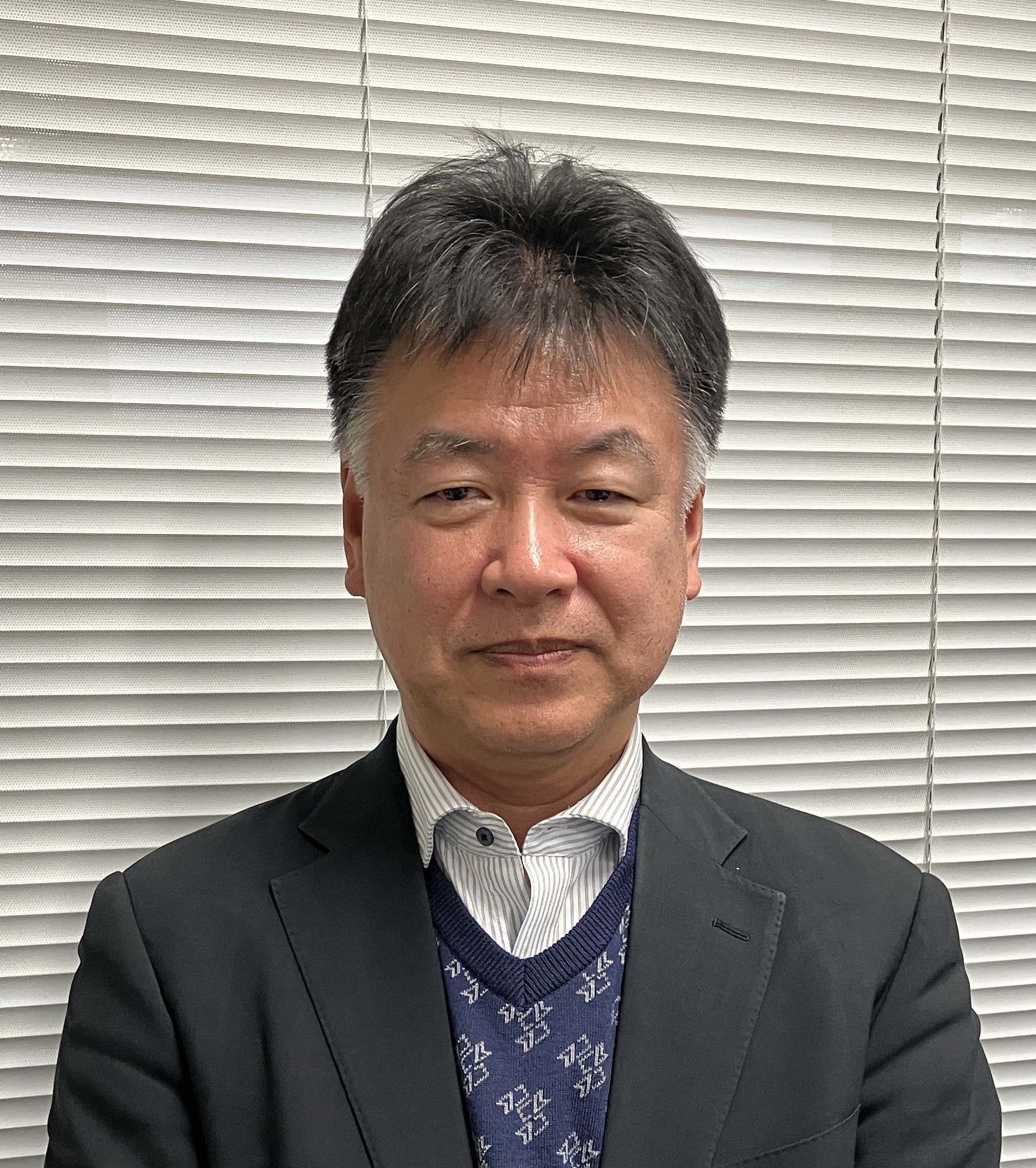}}]{Hiroyuki Kayano}
received his Ph.D. from Saitama University, Saitama, Japan, in 2005. In 1992, he joined Toshiba Corporation, where he has been engaged in research and development of radio frequency and microwave filters. In 2024, he joined the Global Research and Development Center for Business by Quantum-AI Technology (G-QuAT) at the National Institute of Advanced Industrial Science and Technology (AIST) as a Principal Research Manager.
\end{IEEEbiography}

\begin{IEEEbiography}[{\includegraphics[width=1in,height=1.25in,clip,keepaspectratio]{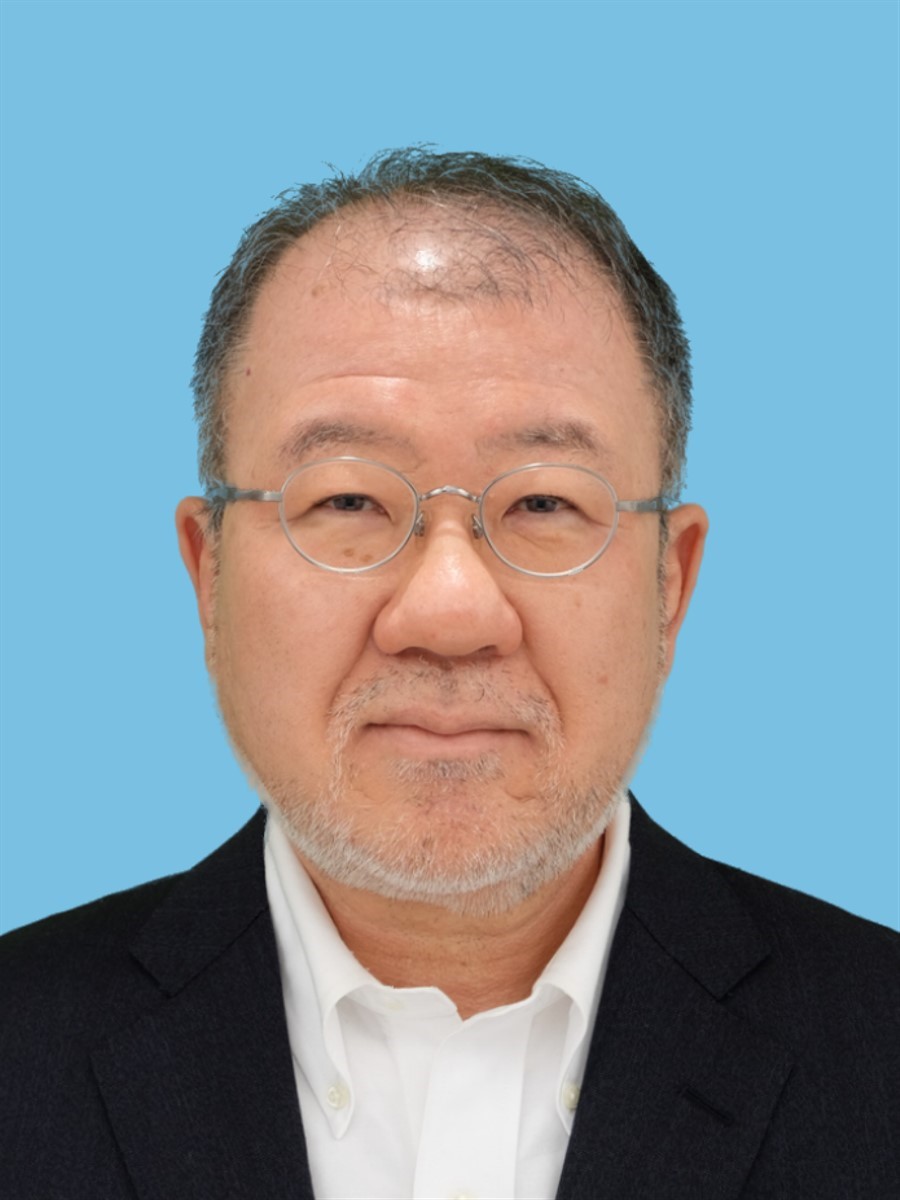}}]{Noriyoshi Hashimoto}
received his bachelor's degree in Electronic Engineering from Tokushima University in 1985 and joined Yokogawa Hewlett-Packard (now Keysight Technologies) the same year. After retiring from Keysight in 2021, he joined the Global Research and Development Center for Business by Quantum-AI Technology (G-QuAT) at the National Institute of Advanced Industrial Science and Technology (AIST) in 2024.
\end{IEEEbiography}

\begin{IEEEbiography}[{\includegraphics[width=1in,height=1.25in,clip,keepaspectratio]{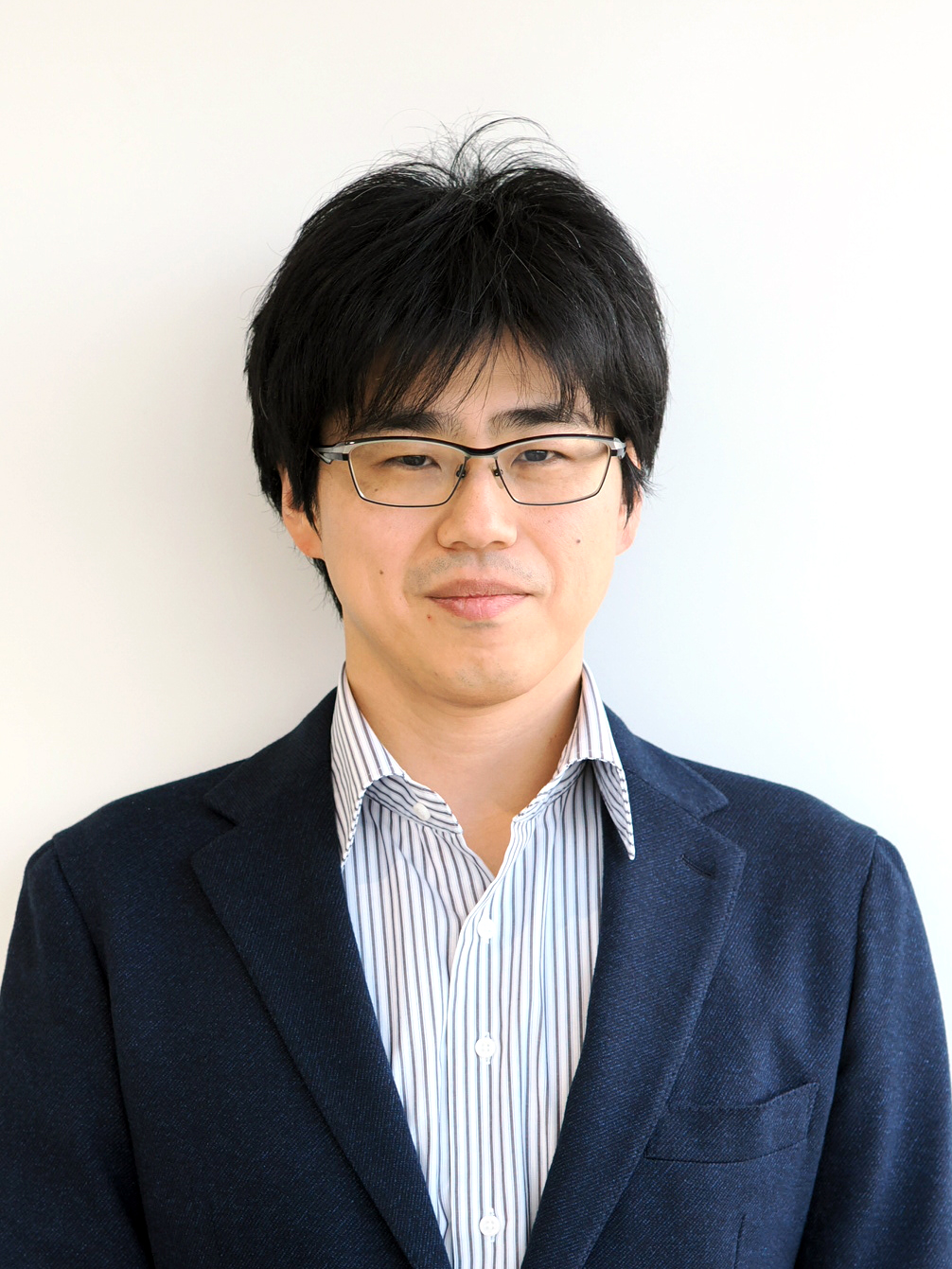}}]{Makoto Minohara}
received his Ph.D. from University of Tokyo, Tokyo, Japan, in 2009. He is currently a Team Leader at the Global Research and Development Center for Business by Quantum-AI Technology (G-QuAT) at the National Institute of Advanced Industrial Science and Technology (AIST). His expertise includes spectroscopy utilizing synchrotron radiation, and his current research interests focus on the development of functional oxide materials, microwave components and cryogenic systems.
\end{IEEEbiography}

\begin{IEEEbiography}[{\includegraphics[width=1in,height=1.25in,clip,keepaspectratio]{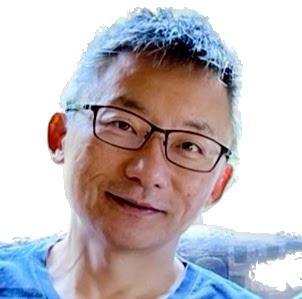}}]{Nobu-Hisa Kaneko}
received his Ph.D. in condensed matter physics from Tohoku University in 1997. After postdoctoral research at Stanford University and SLAC, he joined the National Metrology Institute of Japan (NMIJ) at the National Institute of Advanced Industrial Science and Technology (AIST), in 2003. His work focuses on quantum electrical standards, including the quantum Hall effect and Josephson effect. Currently a Prime Senior Researcher, he also contributes to G-QuAT, AIST's quantum research initiative.
\end{IEEEbiography}

\begin{IEEEbiography}[{\includegraphics[width=1in,height=1.25in,clip,keepaspectratio]{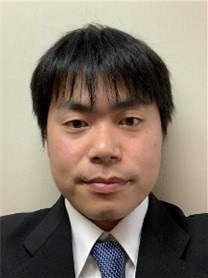}}]{Tomonori Arakawa}
received the B.E. degree in applied physics from Kyoto University, Japan, M.S. degree from Kyoto University in science, Japan and the Ph.D. degree in science from Kyoto University, Japan, in 2009, 2011 and 2014, respectively. Then he became an assistant professor in Graduate School of Science, Osaka University, Japan. He moved to National Institute of Advanced Industrial Science and Technology (AIST), Japan, in 2021, and is currently working as a Senior Researcher in the field of instrumentation at microwave and condensed matter physics.
\end{IEEEbiography}

\vspace{11pt}

\vfill

\end{document}